# Biochars at the molecular level.

# Part 2 – Development of realistic molecular models of biochars.

Rosie Wood[a], Ondřej Mašek[b], Valentina Erastova[a]*

[a] School of Chemistry, University of Edinburgh, Joseph Black Building, David Brewster Road, King's Buildings, Edinburgh, EH9 3FJ, United Kingdom

[b] UK Biochar Research Centre, School of GeoSciences, University of Edinburgh, Crew Building, Alexander Crum Brown Road, King's Buildings, Edinburgh, EH9 3FF, United Kingdom

*valentina.erastova@ed.ac.uk

## Abstract

Biochars have been attracting renewed attention as economical and environmentally friendly carbon sequestration materials with diverse applications. However, experimental developments may be limited by the lack of molecular-level knowledge of the key interactions driving these applications. Molecular modelling techniques, such as molecular dynamics simulations, provide atomistic-level insights into physicochemical processes, informing and guiding experimental development. This paper presents the second part of the work dedicated to unravelling biochars' molecular makeup and developing representative and realistic molecular models for computer simulation. To this end, in Part 1, we reviewed analytical techniques frequently used for the characterisation of biochars and collated a large volume of experimental data. Based on this knowledge and data, in this Part 2, we develop molecular models of three woody biochar materials, representative of real biochars produced under low-, medium-, and high-temperature treatments. We characterise these models, validate them against experimental data, and share our models with the research community. These are the first realistic models of biochars and are shared in an easy-to-use format, with assigned OPLS-AA force field and ready for use in classical molecular dynamics simulations with GROMACS engine. Furthermore, we detail the iterative approach we used for the design of these biochar models, discussing what we have learned about the relationship between biochar composition and morphology thanks to this approach. We also share all the building blocks used to create these biochar models, by this allowing other users to build their custom-made biochars. Through this work, we hope to initiate the uptake of molecular dynamics simulations for the study and development of biochar materials.

# 1. Introduction

Biochars are black carbonaceous solids produced through the pyrolysis of biomass under conditions of little or no oxygen. For centuries, biochar-like materials have been used within agriculture as soil amendments;[1,2] however, more recently, these materials have attracted renewed attention as a means to sequester atmospheric carbon and to replace fossil-based alternatives, and as economical and environmentally-friendly adsorbents with a diverse range of applications.[1,3–5] This has led to an interest in biochars' adsorption properties and the publication of hundreds of experimental studies focused on optimising biochars for specific adsorption-based applications.[6–8] Experimental efforts, however, can often be both time- and resource- consuming, yet, unproductive, slowing progress in the development of this important material. Molecular modelling techniques, such as molecular dynamics (MD) simulations, offer a systematic and reproducible alternative which can both yield atomistic insights into physicochemical processes, such as adsorption, and allow for clear identification of mechanistic trends.[9–11] Molecular modelling could, therefore, assist in the rapid advancement of the development of biochars, just as it has done in areas such as biophysics and drug design. Nevertheless, only a limited number of published works have focused on the molecular modelling of biochars and, in many cases, the models used are vastly oversimplified, with biochars being approximated as pure graphene[12–15] or as small graphitic flakes.[13,15–17] Such models are likely to fail in representing the true nature and properties of these complex materials.

The properties of biochars have been shown to be strongly influenced by the highest treatment temperature (HTT) used during their production.[18] This influence of HTT on commonly measured properties of biochars is now relatively well understood and has been the focus of a number of review papers.[18–23] However, fewer works have been dedicated to the relations between the measured properties and the underlying molecular structures comprising biochars.

In order to create a true representative biochar models, in Part 1 of this work (Wood *et. al.* "Biochars at the molecular level. Part 1 – Insights into the molecular structures within biochars.") we critically reviewed the analytical techniques used to characterise biochars and collected large dataset of experimental measurements for woody biochars, this allowed us to obtain insights into how the molecular structures within biochars vary as a function of HTT.[24] This is Part 2 of the work, where we use this gathered information to develop molecular models representative of woody biochars produced at low (400ºC), medium (600ºC) and high (800ºC) HTTs. We believe that molecular modelling has not yet found use in the biochar research community due to the lack of freely-available molecular models of these materials. Therefore, our models, along with the molecular structures used to generate models and experimental database are freely available via https://github.com/Erastova-group/Biochar_MolecularModels. To ensure transferability and ease of use, our models are set up for simulation in GROMACS,[25,26] one of the fastest and most widely used MD engines, and assigned the OPLS-AA force field,[27,28] which is recognised for its accuracy for condense-phase organic systems.[28–31]

Within this paper, we present both our approach and the resulting molecular models of biochars, ensuring they are representative of woody biochars produced at high, medium and low HTTs. To this end, first, we discuss existing methodologies for the modelling of condensed-phase systems, rationalising the choices made in our work. Second, we outline the iterative procedure we adopted, along with the data used to guide, assess, and validate our molecular models at each step. We also present the key lessons learned throughout the iterative development of biochar models. And, finally, we share our biochar molecular models, characterising them against experimental parameters. Our goal is not only to share the developed models with a wider community but also to provide insights into the molecular structures, which are core to the properties of biochars and biochar-like materials. We hope that through this shared knowledge, other researchers in the field may be able to build upon and further our work.

### 1.1. Molecular dynamics simulations of condensed-phase carbonaceous systems

Classical MD simulations perform Newtonian mechanics on preassigned molecular structures. These molecular structures are described by a set of parameters – known as a force field – which define their bonded and non-bonded interactions. These predefined parameters remain fixed throughout a simulation, meaning chemical bonds cannot be made or broken. The molecular structures input into the simulation must, therefore, be chosen as representative of the system of interest. While many force fields exist for organic compounds[27,28,32–37] and, in principle, their assignment to the structure can be done manually, specific tools [32–36,38–42] are of immense help. However, when large chemically complex systems are of interest, there are very little routine solutions or tools to aid in this assignment and the process can become highly laborious. This complicated assignment step has, thus far, drastically limited the uptake of MD in the study of many complex materials. The problem then arises in generating well-equilibrated condensed-phase structures with morphologies representative of the target material. A number of approaches can be used to do this, including simulated annealing,[43] in which the molecular structures are packed into a simulation box and allowed to gradually condense into an equilibrated solid through a series of heating and cooling cycles, and aggregation from solvent, in which the molecular structures are packed into a simulation box, partially solvated and allowed to assemble into a condensed cluster.

An alternative approach to tackling the setup and simulation of complex systems is the use of *bond order* or *reactive* force fields,[44–50] which allow for chemical transformations throughout an MD simulation. The advantage of this method is that the same set of parameters can be used to describe many bonding states of a given atom and, as a result, molecular structures emerge throughout the simulation. This removes the need to pre-define the molecular structures that comprise the system, and, instead, one can simply input the atomic composition. During the simulation, the atoms "react", leading to the formation of bonds and the evolution of molecular structures. This process is guided by the type and proximity of neighbouring atoms, simulation box density (i.e. packing), the simulation protocol, and the choice of *reactive* force field. While this method is hugely appealing, the problem arises in the assessment and validation of the resulting molecular structures, with minor variations to the set up or simulation protocol resulting in extreme variations.[51–53] *Reactive* force fields have found use in the study of many carbonaceous materials, such as pyro-carbons, kerogens and graphites.[54,55,64,65,56–63] However, due to the method's sensitivity to changes in the simulation set up and protocol, immense care must be taken to ensure the resulting models are chemically sensible and representative of the target material.[51]

## 2. Approach

### 2.1. Approach to the development of biochar molecular models

Our approach to the development of realistic molecular models of biochars is outlined in **Figure 1**. We began by defining the properties of woody biochars produced at low (400ºC), medium (600ºC) and high (800ºC) HTT using experimental data collected in Part 1 of this work.[24] We used these properties as targets throughout our model development, guiding us towards representative molecular models of biochars. Based on these targets, we created a selection of molecular *building blocks*. For each biochar model, we set up a simulation box containing a selection of molecular *building blocks* – our best guess to match the target chemical characteristics of a system. Following a simulated annealing procedure[43] we condensed our systems into solids, periodic in x-, y- and z- directions and, therefore, representative of the bulk material. After equilibrating each condensed system at room temperature and atmospheric pressure, we analysed the physical properties of our bulk models. Through comparison of simulated systems' properties to the gathered experimental data, we identified potential changes and improvements to be made to improve the fit of the model to the target biochars. We then updated our *building blocks* selection and repeated these steps, through an iterative process, until satisfied that we had obtained molecular models representative of each target biochar. Finally, to create two exposed surfaces, we expanded the simulation box of each bulk model in z-direction, by this creating an unoccupied volume above and below the bulk material, and ran one further MD simulation to equilibrate this new structure. For complete set up and simulation details see **Methods Sections 5.3** and **5.4**, respectively. Lastly, we assessed the physical properties of our exposed-surface models and compared against our target properties to ensure our structures remained representative.

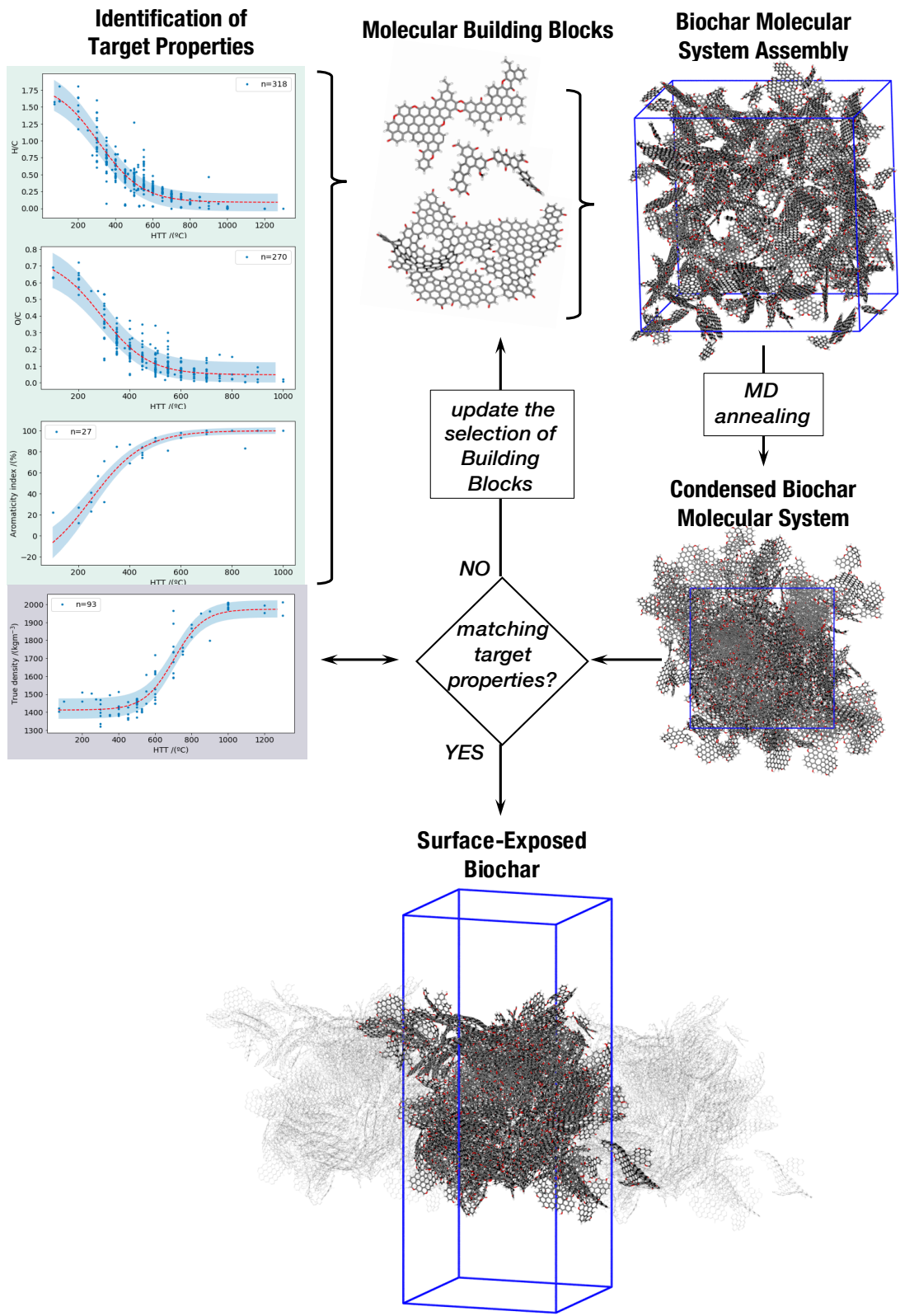

**Figure 1. A schematic guide to the iterative approach used to develop our biochar models.**

## 2.2. Choice of experimental data

In Part 1 of this work, we collected experimental data from literature and used this to gain insights into how the commonly measured properties of biochars vary as a function of HTT.[24] We now use a subset of this data to develop our biochar molecular models. We focus primarily on data that provides clear insights into the chemical compositions or molecular structures within biochars or, which offers quantifiable targets to use throughout our model development. To this end, we selected four key physicochemical properties: H/C and O/C ratios, which describe the chemical compositions of a biochar; *aromaticity indices*, which describe the proportion of aromatic carbons within a biochar; and *true densities*, which describe the density of a biochar at particle-level resolution.[1,66] We also utilised a range of qualitative data, for example, the range of functional groups present within biochars or the morphologies of their molecular structures. A more detailed discussion of these properties can be found in our preceding work.[24]

We chose to create three biochar molecular models, representative of those produced at low (400ºC), medium (600ºC) and high (800ºC) HTTs. We chose these HTTs, as representatives of the common HTT range used in biochar researcher and are, therefore, anticipated to provide relevant and useful insights into the variations within biochars. We also limited ourselves to biochars produced from woody feedstocks due to their relatively low ash content (typically 0 - 10%).[19,67] This choice allows us to neglect ash from our models and focus on the organic components only whilst still creating representative molecular models.

## 2.3. Fitting of experimental data and definition of target biochars

To define the properties of our three biochars, we fit curves to the data – H/C and O/C ratios, *aromaticity indices* and *true density* – using a sigmoidal function (see **Methods Section 5.1** for more details). From these fits, shown on the **Figure 2**, we were able to predict a mean value plus a lower and upper confidence limit, associated to each property at HTTs of 400ºC, 600ºC and 800ºC. This gave us a set of quantifiable target properties to use in the development of each of our biochar molecular models.

Based upon our knowledge of the thermal stabilities of common oxygen-based functional groups in the biochars[24] and the data from a number of published studies,[19,67–77] we estimated the ranges of functional groups present within each biochar-type. The five properties used to guide the biochar model development are summarised **Table 1**.

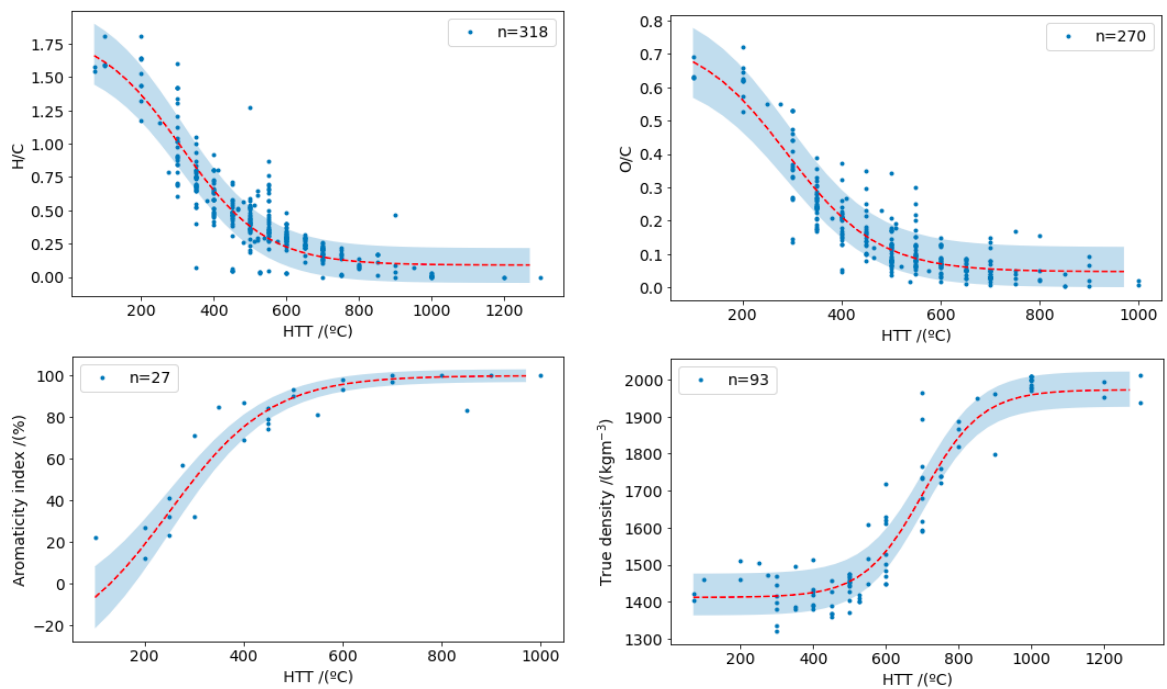

**Figure 2. Changes in the experimentally measurable properties (H/C and O/C molar ratios, *aromaticity index*, *true density*) of woody biochars with the highest treatment temperature. Experimental data is shown as blue points, fitted curve as dashed red line and confidence interval (95%) as pale blue area.**

**Table 1. Chemical and physical properties of woody biochars produced at highest treatment temperatures (HTTs) of 400ºC, 600ºC and 800ºC. Values in brackets are confidence limits.**

| HTT /°C | Chemical properties | | | | Physical properties |
|---|---|---|---|---|---|
| | H/C | O/C | *Aromaticity index* /% | Range of possible functional groups | *True density* /kg m$^{-3}$ |
| 400 | 0.65 (0.49-0.82) | 0.21 (0.15-0.29) | 75 (70-81) | Alkane, alkene and aromatic C, plus wide range of O-based functionalities | 1430 (1380-1490) |
| 600 | 0.23 (0.08-0.36) | 0.07 (0.02-0.15) | 96 (92-99) | Predominantly aromatic C, plus stabilised O-based functionalities | 1540 (1490-1600) |
| 800 | 0.12 (0.00-0.24) | 0.05 (0.00-0.12) | 99 (96-100) | Predominantly aromatic C, plus limited number of aromatic stabilised O-based functionalities | 1850 (1800-1900) |

## 2.4. Development of molecular *building blocks*

Our biochar models are formed of collections of individual molecular *building blocks*. We used the chemical properties (H/C and O/C ratios, *aromaticity indices* and functional groups), outlined in **Table 1,** to develop these structures. Each molecular *building block* consists of a polyaromatic *core*, which makes up its bulk, and is decorated with functionalised *arm groups*, which were added to adjust the *aromaticity indices* and H/C ratios. Alkyl *arm groups* both decrease *aromaticity index* and increase H/C ratio, while aryl *arm groups* increase H/C ratio whilst retaining the high *aromaticity index* of the *core*. Thus, the different properties of each *building block* can be tuned to match those targeted. Molecular structures built in this way are generally referred to as being of *island*-type architecture. They have been found to be present in crude oil components, such as asphaltenes,[78,79] and are commonly used in the molecular modelling of carbonaceous solids.[31,78,80–85] To achieve the desired O/C ratios, each *building block* also features a variety of oxygen- containing functional groups, representative of those present at each HTT. Three example building blocks, including their chemical descriptors, are shown on the **Figure 3.** The Mol IDs correspond to the IDs of these building blocks shared in the https://github.com/Erastova-group/Biochar_MolecularModels sub-directory '/building_blocks', while the chemical descriptors are also available for each of molecular building blocks.

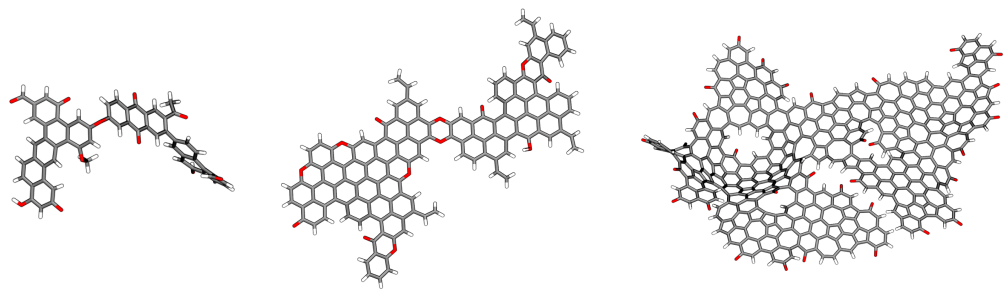

| Mol ID | 0012 | 0020 | 0038 |
|---|---|---|---|
| Formula | $C_{59}H_{32}O_{11}$ | $C_{135}H_{50}O_{13}$ | $C_{429}H_{72}O_{31}$ |
| H/C | 0.54 | 0.37 | 0.17 |
| O/C | 0.19 | 0.096 | 0.072 |
| MW (g/mol) | 916.9 | 1870.9 | 5721.3 |
| Domain Size | 12 | 41 | 155 |
| Aromaticity (%) | 75 | 93 | 100 |
| -C-OH | 2 | 1 | 0 |
| -C=O | 7 | 5 | 31 |
| -C-O-C- | 2 | 7 | 0 |

Figure 3. Visualisations of example building blocks and their chemical descriptors. The Mol ID corresponds to the ID in the building block database. For the renderings carbon atoms are shown in grey, oxygen atoms – in red, hydrogen atoms – in white.

## 3. Results

### 3.1. Key lessons learned during model development

By adopting an iterative approach to the development of our biochar molecular models, we were able to identify structure-property relationships and determine the effects of simulation protocol on the final systems structure and properties. The large confidence intervals (**Table 1**) gave us freedom to explore a relatively large region of chemical space when creating our *building blocks* and, through combinations of these, we constructed our biochar molecular models. We found that different combinations of *building blocks* resulted in different biochar morphologies, and, through numerous iterations, we learned a number of key lessons relevant to the modelling of biochars and biochar-like materials.

From our exploration of chemical space, we were able to observe how the molecular properties of our *building blocks* influenced the structures of our condensed-phase systems. Firstly, we noted the strong relationship between the *aromatic domain size* of our *building blocks* and the *true density* of resulting condensed-phase systems. With larger *aromatic domain sizes* leading to higher *true densities* of the condensed-phase systems. We quantified this relationship by simulating the condensation of a series of graphitic flakes with *aromatic domain sizes* ranging from 1 ring (i.e. benzene) to 272 rings. We identified a logarithmic relationship between the two properties and, from this, we were able to tune the *true densities* of our condensed-phase systems through the *aromatic domain sizes* of our *building blocks*. This is discussed in more detail in **Section S2.1** of the **Supporting Information**. Secondly, we learned that curvature, introduced by non-hexagonal rings, is key in the developing amorphous molecular morphologies such as those seen in non-graphitising materials. This curvature acts to disrupt the packing of molecular *building blocks*, thereby hindering the formation of graphitic stacks or *crystallites*. However, we found this disruption to packing also resulted in the formation of condensed-phase systems with lower *true densities*. Further details on this can be found in **Section S2.2** of the **Supporting Information** and will be discussed later in this work in **Section 3.2.2**.

We also studied the influences of the simulation protocol on the morphology of our final material. As biochar is a condensed material, comprising large molecular structures, the assembly of these structures into a state representative of an experimentally observed one, whilst ensuring ergodicity, is essential. To this end, we investigated several simulation protocols, including the use of aggregation from solvent (**Section S3.1** of the **Supporting Information**) and simulated annealing, using a range of annealing temperatures, annealing steps and pressures (**Sections S3.2** and **S3.3** of the **Supporting Information**). From these investigations, we concluded that annealing was a more favourable approach due to the computational cost of simulating large solvated systems, and that (i) for larger systems, higher annealing temperatures were required in order to better sample the phase-space before condensing our systems, and (ii) higher applied pressures encourages the faster contraction of the simulation box, preventing the formation of individual separated clusters.

Finally, we learned the importance of validating our structures against a wide variety of experimentally measurable properties, including both numerically quantifiable properties, such as *true density*, and properties which are less easy to quantify, such as molecular morphologies from High Resolution Transmission Electron Microscopy (HRTEM) imaging.

### 3.2. Construction of bulk biochar molecular models

### 3.2.1 Construction of bulk models from molecular *building blocks* featuring only hexagonal rings

We first present three bulk models – BC400, BC600 and BC800 – representative of woody biochars produced at 400ºC, 600ºC and 800ºC, respectively, and created with molecular *building blocks* containing no non-hexagonal rings. Visualisations of BC400, BC600 and BC800 are shown in **Figure 4a** and their properties are given in **Table 2**.

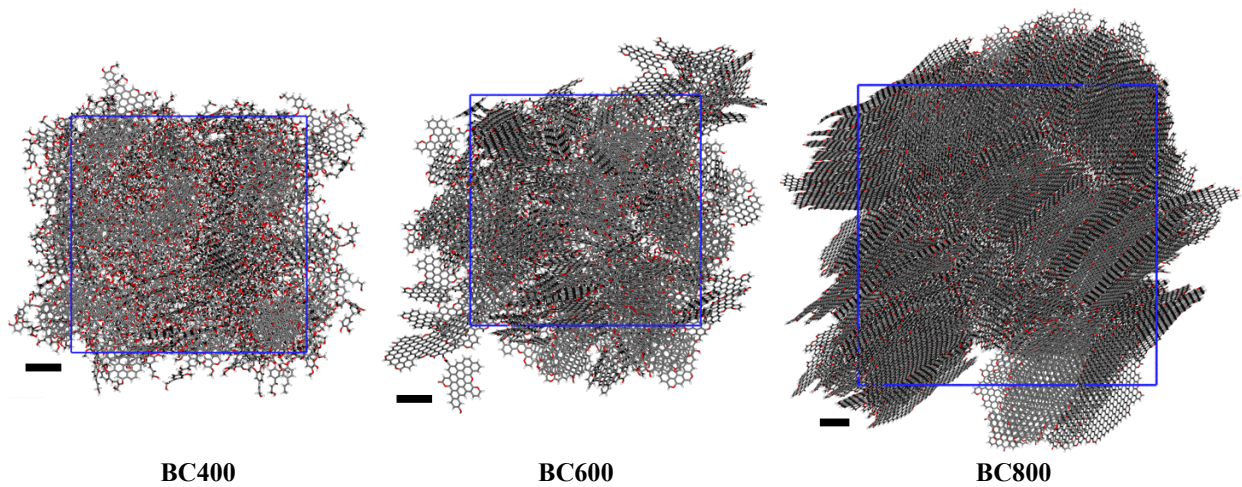

**BC400**   **BC600**   **BC800**

a) Bulk biochar models, built with hexagonal rings only.

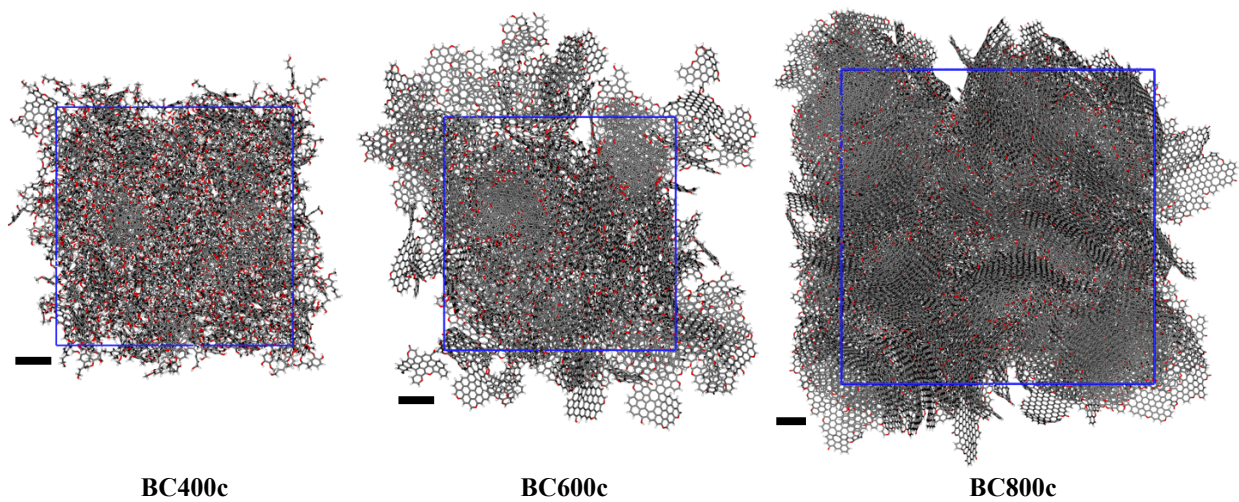

**BC400c**   **BC600c**   **BC800c**

b) Bulk biochar models, built with a proportion of non-hexagonal rings.

**Figure 4. Visualisations of bulk biochar models. Carbon atoms are shown in grey, oxygen atoms – in red, hydrogen atoms – in white. The periodic simulation box is given in blue. Scale bars are 1 nm.**

**Table 2. Properties of BC400, BC600 and BC800.** *True densities* are shown as an average of three repeat simulations.

| Model name | H/C | O/C | *Aromaticity index* /% | *core* aromatic domain size /rings | *True density* /kg m$^{-3}$ |
|---|---|---|---|---|---|
| BC400 | 0.63 | 0.19 | 78 | 23 | 1402 ± 1 |
| BC600 | 0.24 | 0.07 | 98 | 75 | 1581 ± 8 |
| BC800 | 0.09 | 0.04 | 100 | 285 | 1878 ± 3 |

The H/C and O/C ratios, *aromaticity indices* and *true densities* of BC400, BC600 and BC800 all fall well within our target confidence intervals, as defined by **Table 1**. Comparison of simulated TEMs of these models to experimental HRTEMs of woody biochars are shown in **Figure 5.** Further HRTEMs of biochars[23,86–93] and similar materials[57,94–107] can be found on **Figure S8** of the **Supporting Information**, and further support our conclusion that these models are good representations of morphology of target biochar materials at the chosen HTTs. Our high temperature biochar model (BC800), however, appears more slightly more ordered than can be seen in some HRTEMs of woody biochars produced at this HTT[89,91] and so should be thought of as only representative of more graphitic regions of these biochar-types.[87,88,90,93] The molecular *building blocks* used to create these models contained no non-hexagonal rings and so this result is not unexpected.

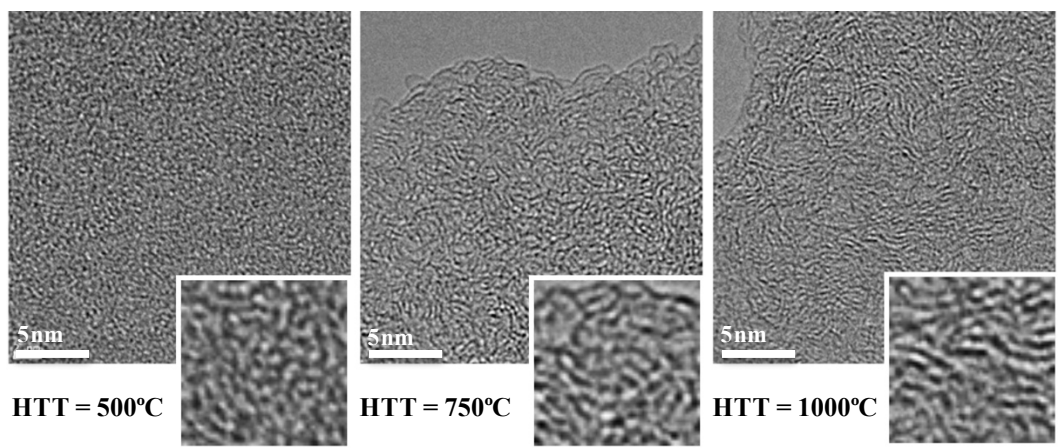

a) Experimental HRTEMs of woody biochars produced at 500ºC, 750ºC and 1000ºC. Reprinted from Carbon, 102, Deldicque, D., Rouzaud, J. N. & Velde, B., *A Raman - HRTEM study of the carbonization of wood: A new Raman-based paleothermometer dedicated to archaeometry*, 319–329, Copyright (2016), with permission from Elsevier.

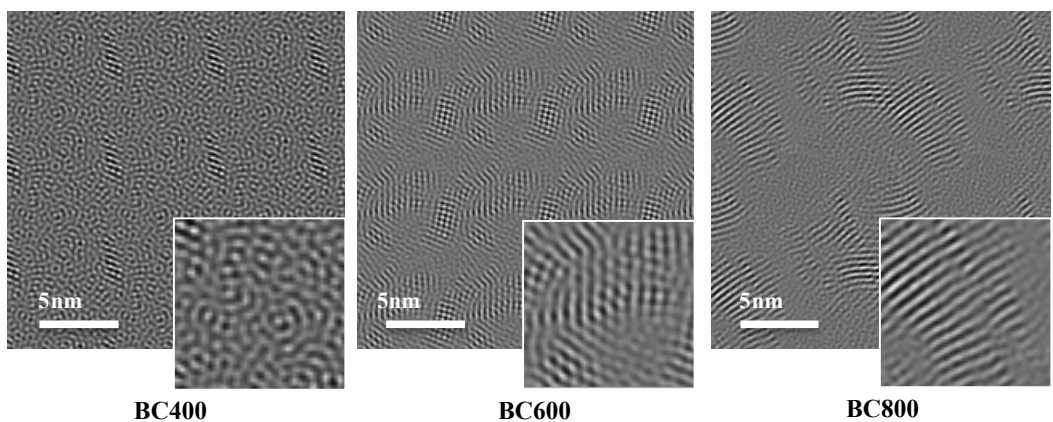

b) Simulated TEMs of BC400, BC600 and BC800 bulk models.

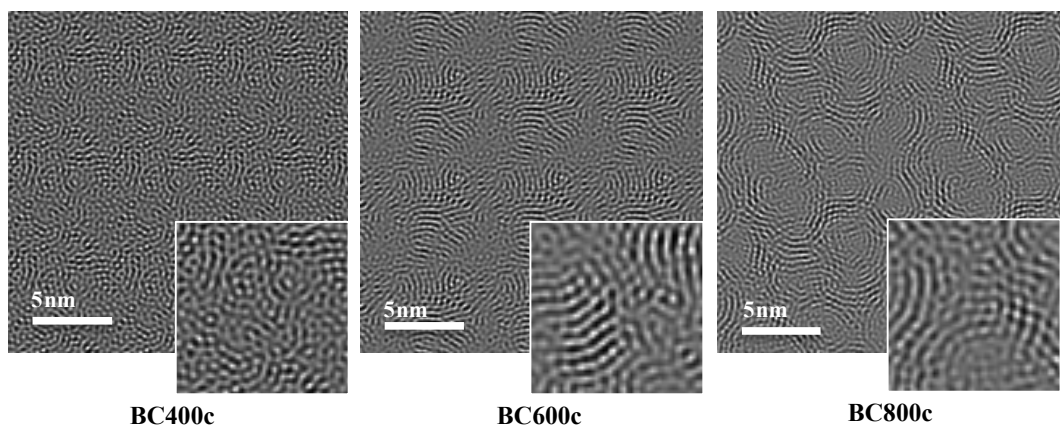

c) Simulated TEMs of BC400c, BC600c and BC800c bulk models.

**Figure 5.** Experimental and simulated TEMs of bulk biochar models. All scale bars are 5 nm and insets are 5×5 nm area.

### 3.2.2 Construction bulk models from molecular *building blocks* containing non-hexagonal rings

To illustrate the effects of including curvature into our molecular *building blocks,* we also developed three bulk models – BC400c, BC600c and BC800c – representative of woody biochars produced at 400ºC, 600ºC and 800ºC, respectively. While these models feature H/C and O/C ratios, *aromaticity indices* and *aromatic domain sizes* comparable to those of BC400, BC600 and BC800, their *building blocks* also contain approximately one pentagonal ring per five hexagonal rings and one heptagonal ring per ten hexagonal rings. These non-hexagonal rings add curvature to our *building blocks* and result in bulk models with distinct structural properties. Visualisations of BC400c, BC600c and BC800c are shown in **Figure 4b** and their properties are given in **Table 3**.

**Table 3. Properties of BC400c, BC600c and BC800c. Values in brackets show percentage difference versus hexagonal-ring only counterparts.**

| Model name | H/C | O/C | *Aromaticity index* /% | *core aromatic domain size* /rings | *True density* /kg m$^{-3}$ |
|---|---|---|---|---|---|
| BC400c | 0.63 | 0.2 | 78 | 23 | 1353 (-3%) |
| BC600c | 0.23 | 0.08 | 98 | 73 | 1496 (-5%) |
| BC800c | 0.10 | 0.04 | 100 | 271 | 1537 (-18%) |

The *true densities* of BC400c, BC600c and BC800c are all decreased versus their more graphitic counterparts (BC400, BC600 and BC800), with those of BC400c and BC800c falling below the targets (**Table 1**) by 2% and 15%, respectively. This is expected, as the curvature within the updated *building blocks* results in less efficient packing and, therefore, lower *true densities*. While matching density is desirable, one should also keep in mind that this is a bulk average property and may vary between sampled areas, being lower at the less packed interface. Furthermore, just as any methods, true density measurements have their limitations, often overestimating the values. (See Part 1 of this work for the detailed discussion of methodologies.[24]) Therefore, we also validate the models through comparison of the simulated TEM, shown in **Figure 5c**, with those of our previous models, shown in **Figure 5b**, also reveals that BC400c, BC600c and BC800c are less ordered than BC400, BC600 and BC800. Again, this is caused by the curvature of the updated *building blocks*, which limits the formation of graphitic stacks and *crystallites*. When compared to experimental HRTEMs (**Figure 5a**), we see that our updated models, containing non-hexagonal rings, better represent the amorphous regions of woody biochars. It is, therefore, reasonable to conclude that non-hexagonal rings should indeed be included in our molecular models. However, further investigations should assess the relationship between of morphology and *true density*.

### 3.3. Construction of biochar molecular models with exposed surfaces

To ensure these can be used to study processes at the interface, we created systems featuring two exposed surfaces from each of our bulk models. We did this by expanding each simulation box in the z-direction, to create a space both above and below the biochar. We then performed a an equilibration simulation allowing the newly created surface molecules to adapt to their new environment. Our molecular models with exposed surfaces are referred to as *BCX*-surf, where *BCX* is the bulk biochar model from which each surface-exposed model was created. Visualisations of these models are shown in **Figure 6**.

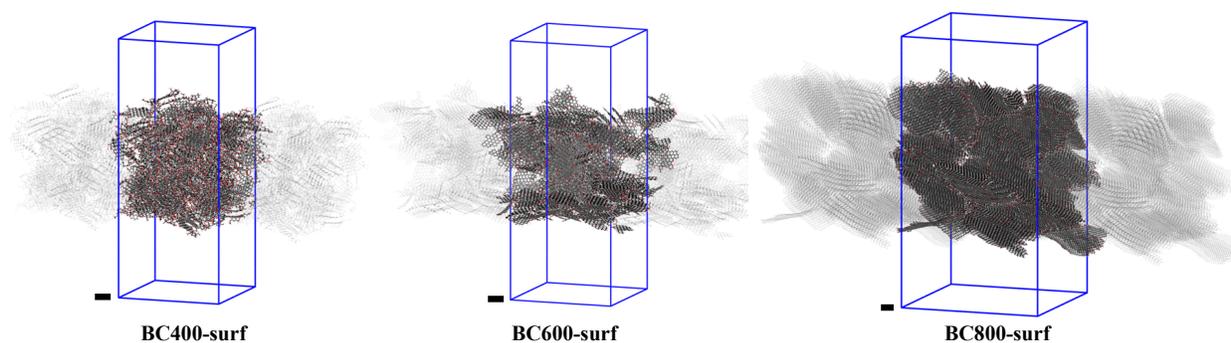

a) Visualisation of surface-exposed models for BC400, BC600 and BC800.

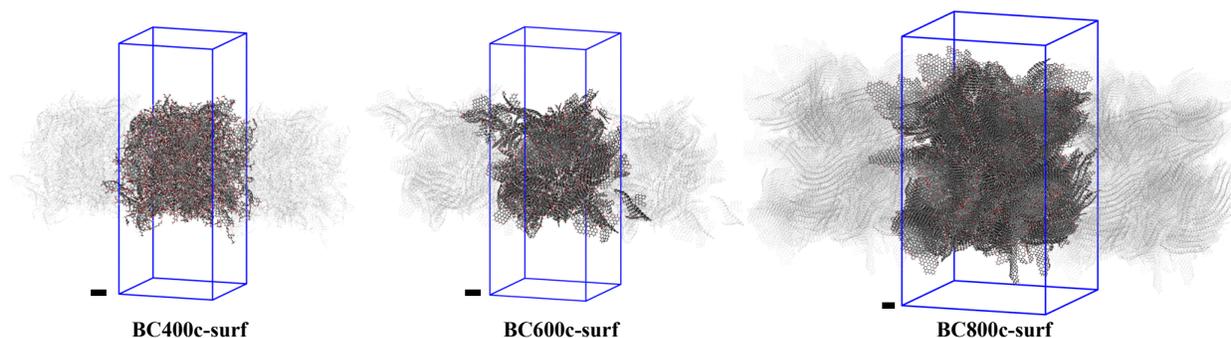

b) Visualisation of surface-exposed models for BC400c, BC600c and BC800c.

**Figure 6. Visualisation of surface exposed models of biochars.** Carbon atoms are shown in grey, oxygen atoms – in red and hydrogen atoms – in white. The periodic simulation box is given in blue, periodic images forming the layer are shown in transparency, scale bars are 1 nm.

In order to quantify the roughness of each surface, we can deploy the solvent accessible surface areas (SASAs). Nevertheless, due to variability of the simulation sizes the absolute values cannot be cross-compared. To this end, we report the normalises SASA per cross-sectional unit area (see **Methods Section 5.5.2**) – an entirely flat surface would return a normalised SASA value of 1, whereas rough surfaces would have proportionally larger normalised SASAs. The normalised SASAs for each surface-exposed system are given in **Table 3.**

All values appear significantly higher than 1, indicating a high degree of surface roughness. Our normalised SASAs also increase with increasing HTTs, and are higher for the structures involving curved *building blocks*. This same trend is seen in laboratory produced biochars, where *specific surface area* (SSAs) and porosity also increase with increasing HTT.[72,76,108–114]

Table 3. Normalised specific surface areas of the biochar surface-exposed systems.

| Model name | Normalised SASA per cross-section area (unitless) |
|---|---|
| BC400-surf | 1.62 |
| BC600-surf | 1.97 |
| BC800-surf | 2.90 |
| BC400c-surf | 1.96 |
| BC600c-surf | 2.16 |
| BC800c-surf | 4.42 |

The goal of this work is to develop biochar surface for study of the interfacial phenomenon, such as pollutant adsorption, gas sequestration or surface wetting. Molecular models of biochar materials, developed in this work, are of a significant system size, which can make their simulation a computationally demanding task, even when using most advanced software and high-performance computing resources. Bulk material is of limited importance when modelling the interfacial properties and surface interactions: close to the surface, bulk material acts as structural support for surface-exposed atoms, but beyond a few nanometres thickness, it becomes redundant. While excluding unnecessary material can save significant amounts of computational resources.

To maximise the area of surface sampled whilst minimising computational expense, our models are created with very high surface-to-mass ratios and with small amount of bulk material. As a result, the *specific surface areas* (SSAs) of our surface-exposed biochar models should not be directly compared to those found within literature. Furthermore, experimental measurements of SSAs of biochar include a wide range of pore sizes, from sub-nanometre pores to tens of micrometres, yet our models feature microporosity, in limited amounts. Future work will focus on including porosity into our molecular models, to better represent the range of pore sizes present within biochars.

We anticipate, that one of the main uses of our surface-exposed biochar models will be to study adsorption onto biochars using MD simulation, and we recognise that experimental adsorption data is often reported on a *per mass* basis. To compare simulated adsorption data with experimental adsorption data, we recommend expressing adsorption of both computationally acquired data and experimental studies in *per surface* units. This can easily be achieved when SSA of an adsorbent is reported alongside its experimental adsorption data.

## 4. Conclusion

In this paper, we share our work on the development of molecular models representative of woody biochars produced at low (400ºC), medium (600ºC) and high (800ºC) highest treatment temperatures. We first presented a set of three bulk models – BC400, BC600 and BC800 – created using molecular *building blocks* containing only hexagonal rings. These models replicate a wide variety of chemical and physical properties of our target biochars. However, they consist of relatively ordered structures which do not represent the amorphous regions of these materials. These models should, therefore, be thought of as only representative of more graphitic or *crystallite*-containing regions of our target biochars. We then present a second set of three bulk models – BC400c, BC600c and BC800c – created using molecular *building blocks* updated to also contain a proportion of non-hexagonal rings, which add curvature into our molecular structures. The morphologies of these molecular models better represent the amorphous regions of our target biochars, yet we found that their *true densities* were adversely affected by the inclusion of curvature into our structures. From these six bulk models we created surface-exposed systems, which can be used in a wide range of MD-based studies of interactions at the biochar surfaces, for example adsorption studies. All our models presented in this work, along with the experimental dataset used to guide their development, are available for download via our GitHub page (https://github.com/Erastova-group/Biochar_MolecularModels).

We anticipate that this work will not only facilitate the advancement of biochars as adsorbents in contaminant management, through better understanding of the influence of preparation and composition on biochars structure and adsorption properties. We hope this work become a stepping-stone to encourage the uptake of molecular modelling within the biochar research community as a whole.

**Acknowledgements**

Rosie Wood would like to thank E4 DTP for the funding of the PhD project "Molecular Modelling for Design of Biochar Materials". Valentina Erastova would like to thank Chancellor's Fellowship by the University of Edinburgh. The simulations have been performed on Cirrus, an EPSRC-funded Tier-2 high performance computing facility.

## TOC

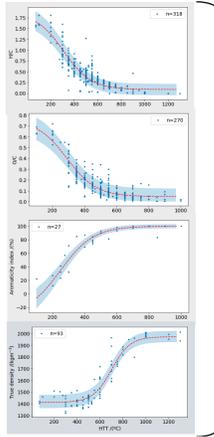
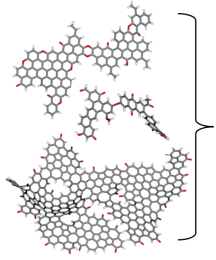
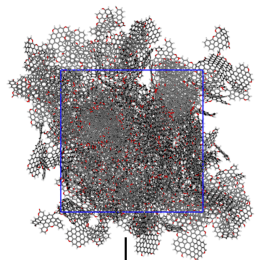
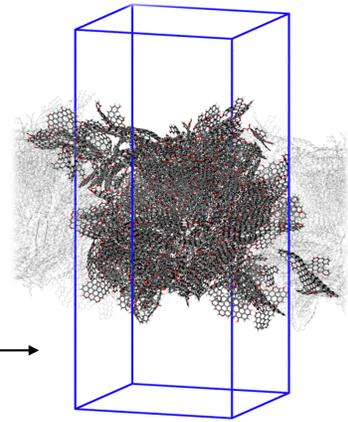
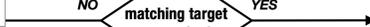

**Identification of Target Properties** | **Molecular Building Blocks** | **Bulk Biochar Molecular System** | **Surface-Exposed Biochar**

## 5. Methods

### 5.1 Fitting of experimental datasets

We collected literature data on woody biochars produced across a range of HTTs. For each property, we aggregated values by HTT and identified mean, lower and upper quantiles (0.025 and 0.975 respectively) using the Python libraries Pandas[115,116] and NumPy.[117] We then used the curve fitting tool from the Python library SciPy[118] to fit curves to our mean datasets using a sigmoidal function (**Equation 1**). This gave us four optimised fit parameters for each averaged dataset: $L$, the height of the curve, $k$, a parameter associated to the slope of the curve, $x_0$, the midpoint of the curve, and $b$, the y-intercept.

$$y = \frac{L}{1+\exp(-k\,(x-x_0))} + b \quad \textbf{(Equation 1)}$$

We then, for each dataset, held the values of , $k$ and $x_0$ at their optimised values and fit two further sigmoidal curves to the lower and upper quantiles within each dataset. This meant the shape of the curve was held constant whilst allowing for transformations in the y-direction.

Our optimised parameters are given in **Tables S1-3**. We used these to calculate the average value each property would take at each of our chosen HTTs, along with lower and upper confidence limits.

### 5.2 Choice of MD engine and force field

We carried out all simulations using GROMACS 2021 or 2022[25,26] and chose the OPLS-AA[27,28] (2001) force field as a base for our simulations of biochars. This is an all atom force field which has been shown to be effective in reproducing the properties of organic liquids.[28–31]

### 5.3 Construction of biochar molecular models

#### 5.3.1 Construction of molecular *building blocks*

We used the properties outlined in **Table 1** to develop molecular *building blocks* representative of each biochar-type. We manually constructed these *building blocks* using Marvin Sketch then used LigParGen[38–40] and PolyParGen[41,42] (OPLS parameterisation tools) to assign OPLS-AA parameters to each.

We used the Python library RDKit[119] to calculate the properties of each of these molecular *building blocks* (number of atoms, chemical formula, molar ratios, molecular weight, number of aromatic carbons, number of conjugated aromatic rings, number of functional groups) and calculated *aromaticity indices* using:

$$Aromaticity\ Index\ (\%) = \frac{C_A}{C_T} \times 100 \qquad \text{(Equation 2)}$$

where $C_A$ is the number of aromatic carbons and $C_T$ is the total number of carbons in each molecular *building block*.

We ran energy minimisations on all molecular *building blocks* prior to further use (details in **Section 5.4.1)**

**5.3.2 Condensation of molecular *building blocks* and creation of bulk biochar molecular models**

For each biochar model, we placed molecular *building blocks* into a cubic simulation box and ran an energy minimisation using the procedure detailed in **Section 5.4.1.**

We then simulated the condensation of each system. We began with 10 ns equilibration in the canonical ensemble (NVT), following the procedure outline in **Section 5.4.2**. The reference temperatures and timesteps that we used during this first equilibration are detailed in **Table 4.**

**Table 4. NVT simulation set up.**

| Model name | Reference temperature /K | Timestep /fs |
|---|---|---|
| BC400 | 1000 | 1 |
| BC600 | 2000 | 0.5 |
| BC800 | 3000 | 0.5 |

We then gradually cooled our systems throughout a 25 ns simulation in the isothermal-isobaric ensemble (NPT), following the procedure outlined in **Section 5.4.3** and using isotropic pressure coupling. The reference temperatures and timesteps that we used during this second equilibration are detailed in **Table 5**.

**Table 5. NPT annealing simulation set up, where T is temperature and *t* is time**

| System | Ref T at $t = 0$ ns /K | Ref T at $t = 10$ ns /K | Ref T at $t = 20$ ns /K | Ref T at $t = 25$ ns /K | Total simulation time /ns | Timestep /fs |
|---|---|---|---|---|---|---|
| BC400 | 1000 | 1000 | 300 | 300 | 25 | 1 |
| BC600 | 2000 | 2000 | 300 | 300 | 25 | 0.5 |
| BC800 | 3000 | 3000 | 300 | 300 | 25 | 0.5 |

Finally, we carried out one further 10 ns equilibration in the isothermal-isobaric ensemble (NPT), following the procedure outlined in **Section 5.4.3** and using isotropic pressure coupling.

Following annealing of each system, we first checked our simulations had no artefacts, such as large voids, or long needle-like crystals. We then calculated the densities of each system using the procedure outlined in **Section 5.5.1** and generated simulated TEMs using the procedure outlined in **Section 5.7**. We also repeated our simulations three times, using different starting configurations, to ensure the ergodicity.

### 5.3.3 Creation of biochar molecular models with exposed surfaces

To create an exposed surfaces on each of our biochar models (and repeats), we elongated their simulation boxes in the z-direction such that each biochar layer was separated by approximately 10 nm of space. The system was then energy minimised and equilibrated over 10 ns, following the procedure outlined in **Section 5.4.1** and **Section 5.4.4**, respectively. We used semi-isotropic pressure coupling during this equilibration, thereby allowing the surface-forming xy plane to be decoupled from the interlayer z- direction.

Following this equilibration, we first checked our systems for artefacts, then calculated solvent accessible surface areas (SASAs) using the last 2 ns of the simulation. The SASA reported is for the given system size, therefore, we normalised SASAs of each system using the procedure outlined in **Section 5.5.2**.

## 5.4 Simulation details

We based our simulation procedure on the methodologies outlined by Ungerer *et al.* in their simulations of kerogens.[43]

### 5.4.1 Energy minimisations

Energy minimisations were carried out using the steepest descents algorithm, and proceeded until, on each atom, the maximum force felt was less than 500 kJ mol$^{-1}$ nm$^{-1}$. Each energy minimisation used periodic boundary conditions in x, y and z, a 1.4 nm Van der Waals cut-off and Particle-Mesh-Ewald (PME) electrostatics.

### 5.4.2 Equilibration in canonical ensemble (NVT)

For our simulations in the canonical ensemble, we used periodic boundary conditions in x, y and z, PME electrostatics, a 1 nm Van der Waals cut-off and a Velocity-rescale thermostat with a 0.1 ps coupling constant.

### 5.4.3 Simulated annealing in isothermal-isobaric ensemble (NPT)

For our simulated annealing in the isothermal-isobaric ensemble, we used periodic boundary conditions in x, y and z, PME electrostatics, a 1 nm Van der Waals cut-off and a Velocity-rescale thermostat with a 0.1 ps coupling constant. We also applied pressure coupling using a Berendsen barostat with a 1 ps coupling constant and a reference pressure of 100 bar. We cooled our systems using a four-stage simulated annealing procedure – starting at high temperature and gradually cooling to room temperature (300 K).

### 5.4.4 Equilibration in isothermal-isobaric ensemble (NPT)

For our final equilibrations, we used periodic boundary conditions in x, y and z, PME electrostatics, a 1.4 nm Van der Waals cut-off and a Velocity-rescale thermostat with a 0.1 ps coupling constant. We also applied pressure coupling using a Berendsen barostat with a 1 ps coupling constant and reference pressure of 1 bar. For all systems, we used a reference temperature of 300 K and a 2 fs timestep.

### 5.5 Analysis of simulations

We checked our systems had equilibrated using the inbuilt GROMACS *rms* and *energy* analysis tools. We looked for plateaus in the rms, densities and box dimensions to indicate convergence of our structures. If satisfied, we then carried out visual checks of the structures and trajectories.

### 5.5.1 Density calculations

We calculated the densities and free volumes of each system using the GROMACS *freevolume* tool. We used a probe radius of 0.13 nm (representative of He) and averaged over only the converged 300 K portion of our simulations. We then calculated the density of our biochar using:

$$\rho_{solid} = \frac{\rho_{system}}{1 - V f_{0.13}} \quad \text{(Equation 3)}$$

where $\rho_{solid}$ and $\rho_{solid}$ are the densities of the condensed solid and of the system respectively, and $V\,f_{0.13}$ is the free volume of the system using a probe radius of 0.13 nm.

### 5.5.2 Solvent accessible surface area and normalised solvent accessible surface area calculations

On our exposed surface models, we calculated the solvent accessible surface areas (SASAs) using the GROMACS *sasa* program, using probe radius of 0.18 nm (representative of $N_2$ gas). Due to variations in the dimensions of each of our systems, we normalised our SASAs using **Equation 4**, where $nSASA$ is the normalised SASA and $A_{xy}$ is the $xy$ cross sectional area. This allowed for comparison between the different molecular models.

$$nSASA = \frac{SASA}{2 \times A_{xy}} \quad \text{(Equation 4)}$$

### 5.6 Visualisation, rendering and plotting

Simulation box structures were rendered with VMD[120] v.1.9 with carbon atoms in grey, oxygen atoms in red, nitrogen atoms in blue, hydrogen atoms in white, unless otherwise specified. A periodic box is shown in blue and 1 nm scale bar given at the bottom right of each image.

All plots were produced with Python 3.6 using the Python library Matplotlib.[121]

### 5.7 Simulated TEMs

We generated simulated TEMs of our biochar models using the *ctem* tool of *compuTEM*.[122] We used a beam energy of 200 KeV, an aperture of 0.02 radians and defocus of 20 nm.[55,122]

# Biochars at the molecular level.

# Part 2 – Development of realistic molecular models of biochars.

**Rosie Wood[a], Ondřej Mašek[b], Valentina Erastova[a]***

[a] School of Chemistry, University of Edinburgh, Joseph Black Building, David Brewster Road, King's Buildings, Edinburgh, EH9 3FJ, United Kingdom

[b] UK Biochar Research Centre, School of GeoSciences, University of Edinburgh, Crew Building, Alexander Crum Brown Road, King's Buildings, Edinburgh, EH9 3FF, United Kingdom

*valentina.erastova@ed.ac.uk

# Supplementary Information

## S1. Optimised parameters for experimental datasets

We fit **Equation 1** of the **Main Text** to our experimental data sets. For sake of clarity, this equation is reproduced here as **Equations S1**, where $L$ is the height of the curve, $k$ is a parameter associated to the slope of the curve, $x_0$ is the midpoint of the curve and $b$ is the y-intercept.

$$y = \frac{L}{1+exp(-k*(x-x_0))} + b \qquad \text{Equation S1}$$

We first fit the curve to the mean values at each temperature entry of the datasets. To fit define the upper and lower confidence limits, while maintaining the shape of the curve, we kept the fitted values of $k$ and $x_0$ constant and fitted $L$ and $b$ to the values within two standard deviations from the mean at each temperature point. We used Python libraries Pandas[29,30] and NumPy[31] and the curve fitting tool from the Python library SciPy[32].

**Table S1.** Optimised parameters from fitting of Equation 1 of the Main Text to means of experimental datasets. The determination coefficient, $R^2$, is given for each.

| Experimental property | Optimised parameters | | | | $R^2$ |
|---|---|---|---|---|---|
| | $L$ | $k$ | $x_0$ | $b$ | |
| H/C | 1.77 | -0.01 | 310.09 | 0.09 | 0.96 |
| O/C | 0.71 | -0.01 | 289.83 | 0.05 | 0.97 |
| Aromaticity (%) | 131.57 | 0.01 | 249.14 | -31.72 | 0.99 |
| True density /kg m$^{-3}$ | 560.41 | 0.01 | 701.29 | 1412.12 | 0.98 |

**Table S2.** Optimised parameters from fitting of Equation 1 of the Main Text to lower and upper quantiles of experimental datasets. Values of $k$ and $x_0$ were fixed at optimised values detailed Table S1. The determination coefficient, $R^2$, is given for each.

| Experimental property | Fixed values | | Optimised parameters | | $R^2$ |
|---|---|---|---|---|---|
| | $k$ | $x_0$ | $L$ | $b$ | |
| H/C | -0.01 | 310.09 | Lower 1.68 | Lower -0.04 | Lower 0.92 |
| | | | Upper 1.90 | Upper 0.22 | Upper 0.88 |
| O/C | -0.01 | 289.83 | Lower 0.64 | Lower 0.00 | Lower 0.88 |
| | | | Upper 0.74 | Upper 0.12 | Upper 0.90 |
| Aromaticity (%) | 0.01 | 249.14 | Lower 146.37 | Lower -49.48 | Lower 0.98 |
| | | | Upper 117.03 | Upper -14.05 | Upper 0.96 |
| True density /kg m$^{-3}$ | 0.01 | 701.29 | Lower 563.66 | Lower 1363.38 | Lower 0.97 |
| | | | Upper 546.09 | Upper 1476.08 | Upper 0.91 |

## S2. Exploration of chemical space and structure-property relationships

We began our work by exploring chemical space and identifying structure-property relationships within our condensed-phase systems. We utilised a range of molecular *building blocks* throughout this process and were therefore able to discern a number of key structure-property relationships within our condensed-phase structures. We began with highly diverse *building blocks,* which we combined into simulation cells to approximately match our target properties, before gradually moving towards molecular *building blocks* with more targeted designs. A database of these molecular *building blocks* can be found at https://github.com/Erastova-group/database_building_blocks.

### S2.1. Relationship between *aromatic domain size* and *true density* through condensation of graphitic flakes.

From our initial iterations we observed a correlation between the *true densities* of our condensed-phase systems and the *aromatic domain sizes* of our molecular *building blocks*. We therefore decided to investigate this relationship through the condensation of a series of graphitic flakes with a varying *aromatic domain sizes*. To do this, we created constructed a series of unfunctionalised molecular *building blocks* with *aromatic domain sizes* ranging from 1 ring (i.e. benzene) to 272 rings. We then placed these into cubic simulation boxes and condensed our systems, following the procedure outlined in the **Methods Section 5.4.2** of the **Main Text**. We carried out a number of repeats, using different starting configurations, for each different *aromatic domain size*.

For each *aromatic domain size*, we found the mean *true density* of our systems using the Python libraries Pandas[29,30] and NumPy[31]. We then used the curve fitting tool from the Python library SciPy[32] to fit curves to our mean datasets using a logarithmic function (**Equation S2**). This gave us two optimised fit parameters: $k$ and $b$.

$$y = b + k * \ln(x)$$
**Equation S2**

Our data, along with its fitted curve, is shown in **Figure S1** and our optimised fit parameters are given in **Table S3**. Using this fitted curve, we were able to the *aromatic domain sizes* associated to the target *true densities* of each of our biochar-types (**Table 1** of the **Main Text**). These predicted values are shown in **Table S4.** We note, however, that these values are likely to be underestimates as the unfunctionalised molecular *building blocks* used within this inverstgation are significantly less complex than the molecular *building blocks* used to construct our biochar models, whose irregular shapes, *arm* groups and additional functionalities are likely disrupt the packing of our condensed-phase systems, leading to lower *true densities*.

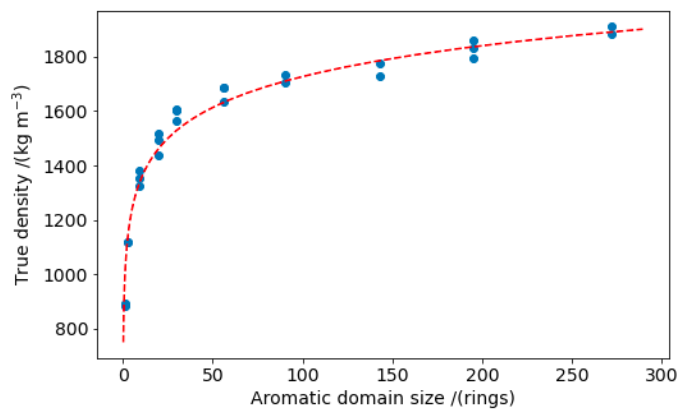

**Figure S1. Relationship between *aromatic domain size* and *true density*. Data shown as blue points and fitted curve as dashed red line.**

**Table S3.** Optimised parameters from fitting of Equation S2 to means of *aromatic domain size* and *true density* data.

| Optimised parameters | | $R^2$ |
|---|---|---|
| *k* | *b* | |
| 163.81 | 971.52 | 0.96 |

**Table S4.** Predicted *aromatic domain size* for each biochar-type.

| HTT /(°C) | Target true density /(kg m$^{-3}$) | Predicted *aromatic domain size* /(number of rings) |
|---|---|---|
| 400 | 1430 | 16 |
| 600 | 1540 | 32 |
| 800 | 1840 | 201 |

### S2.2. Non-hexagonal rings to encourage amorphous structures

Our first collection of molecular *building blocks* comprised only hexagonal aryl rings and carbon- and oxygen-based functionalities. We found that many of the resulting condensed-phase structures exhibited high degrees of ordering due to π-π stacking between flat aromatic sheets within their *building blocks*. An example of this is shown in **Figure S2a**. Although graphitic regions, also known as *crystallites*, are known to be present within biochars, we noted that the particularly high degree of alignment observed in some of these models resulted in simulated TEMs (e.g. **Figure S2b**) which bore strong resemblance to those of graphitising carbons.[1–4] We, therefore, set out to improve the morphologies of our models through inclusion on non-hexagonal rings into our molecular *building blocks*.

Non-hexagonal rings are known to exist in non-graphitising carbons and are thought to bring about the amorphous morphologies within these materials by introducing curvature into the aromatic sheets that comprise them.[1,5–12] Indeed, upon adding these non-hexagonal rings into our molecular *building blocks*, we found that our condensed-phase structures became substantially more disordered. An example of this is shown in **Figure S3a**. Simulated TEMs of our condensed phase structures (e.g. **Figure S3b**) were also substantially improved by the inclusion of non-hexagonal rings and bore greater resemblance to those of non-graphitising materials.[1,2,19,20,3,4,13–18] However, we also noticed that the *true densities* of our condensed-phase systems substantially decreased when non-hexagonal rings were present, likely due to the less space-efficient packing of molecular *building blocks* within these systems. We, therefore, concluded that, whilst non-hexagonal rings encourage the development of amorphous biochar models, more work is needed in order to ensure these models still replicate the target properties.

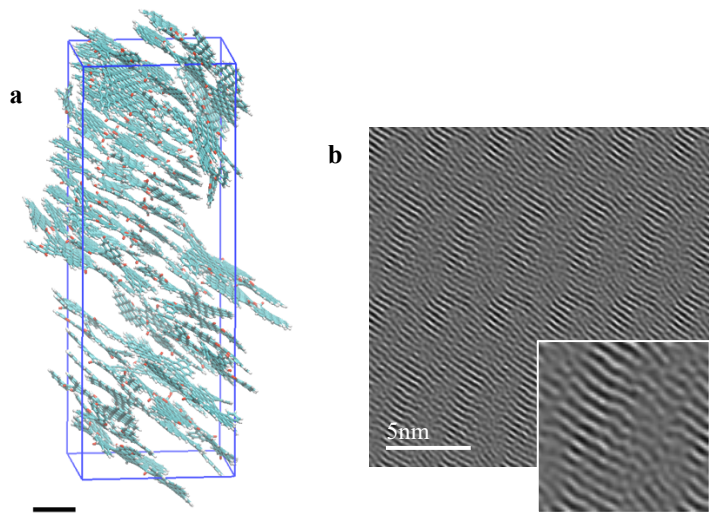

**Figure S2.** Example of a highly ordered system created using *molecular building blocks* containing no non-hexagonal rings. (a) Shows a visualisation of the system with carbon atoms in cyan, oxygen atoms in red and hydrogen atoms in white. The periodic simulation box is given in blue. Scale bar is 1nm. (b) Shows a simulated TEM of the system. Insets is 5x5 nm area.

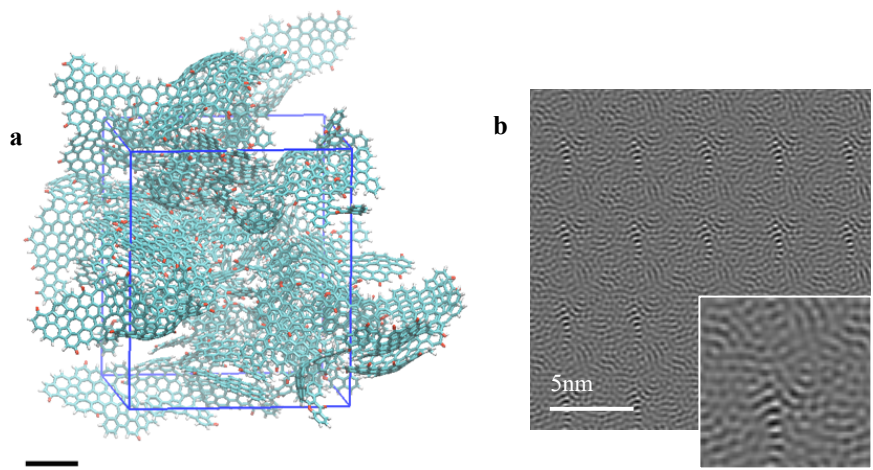

**Figure S3.** Example of a more disordered system created using *molecular building blocks* containing non-hexagonal rings. (a) Shows a visualisation of the system with carbon atoms in cyan, oxygen atoms in red and hydrogen atoms in white. The periodic simulation box is given in blue. Scale bar is 1nm. (b) Shows a simulated TEM of the system. Insets is 5x5 nm area.

## S3. Comparison of simulation protocols

### S3.1. Aggregation from solvent

We briefly explored the idea of using solvent to form our condensed-phase systems.[21–24] We explored a number of different solvents, with the goal of partially solvating our molecular *building blocks* and allowing them to gradually aggregate into solid particles. However, we found that this method came with relatively high computational costs and, therefore, decided not to continue with it.

### S3.2. Anisotropic pressure coupling during annealing

We repeated our simulations of BC400, BC600 and BC800 using anisotropic pressure coupling, thereby allowing the dimensions of our simulation cell to relax independently of one another. Visualisations of these condensed-phase structures – BC400_an, BC600_an and BC800_an – are shown in **Figure S4**, and their properties are given in **Table S5**.

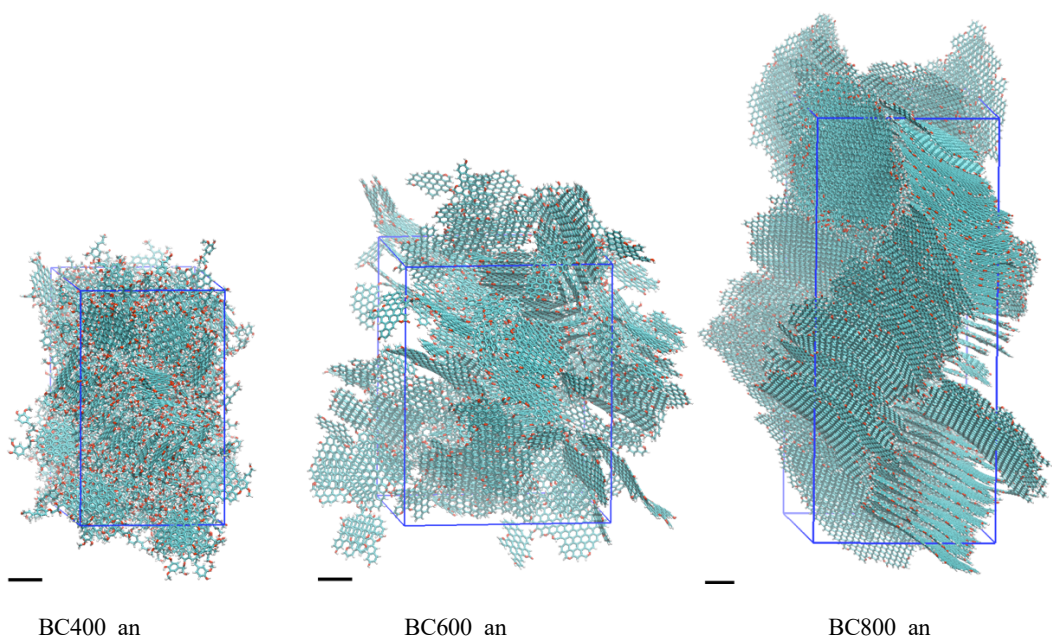

BC400_an    BC600_an    BC800_an

**Figure S4. Visualisations of BC400_an, BC600_an and BC800_an. Carbon atoms are shown in cyan, oxygen atoms are shown in red and hydrogen atoms are shown in white. Scale bars are 1 nm.**

**Table S5. Properties of BC400_an, BC600_an and BC800_an.**

| Model name | H/C | O/C | Aromaticity index /% | *core* aromatic domain size /rings | True density /kg m$^{-3}$ |
|---|---|---|---|---|---|
| BC400_an | 0.63 | 0.19 | 78 | 23 | 1406 |
| BC600_an | 0.24 | 0.07 | 98 | 75 | 1585 |
| BC800_an | 0.09 | 0.04 | 100 | 285 | 1892 |

The *true densities* of BC400_an, BC600_an and BC800_an were comparable to those seen when using isotropic pressure coupling (**Table 2** of the **Main Text**). These results suggest the type of pressure coupling had limited effect on the final *true densities* of our condensed-phase systems and that, instead, molecular composition is central in determining this property.

Simulated TEMs of these models, shown in **Figure S5,** revealed an increase in ordering in both higher temperature models compared to those of their isotropic counterparts (**Figure 4** of the **Main Text**). We also noticed that, when using anisotropic pressure coupling, our simulation cells tended to elongate preferentaily in one direction. These results likely arise from prefential growth of *crystallites* in the direction normal to the aromatic plane. This leads to the formation of stacks of molecular *building blocks*, elongation of the simulation box in the direction normal to the aromatic plane and collapse of the simulation cell in the other two dimensions, resulting in cell dimensions on the order of one molecular length. This behaviour encourages the preferential ordering of the molecular *building blocks*, which would be a physical behaviour in a constraint-free equilibrated system, this may not necessarily be the case of biochar. Furthermore, the collapse of the simulation box in the direction also allowed self-interactions between the *building blocks*. Therefore, having observed that the system densities are consistent between the methods, and to avoid unphysical interactions throughout our simulations, we chose to proceed using isotropic pressure coupling.

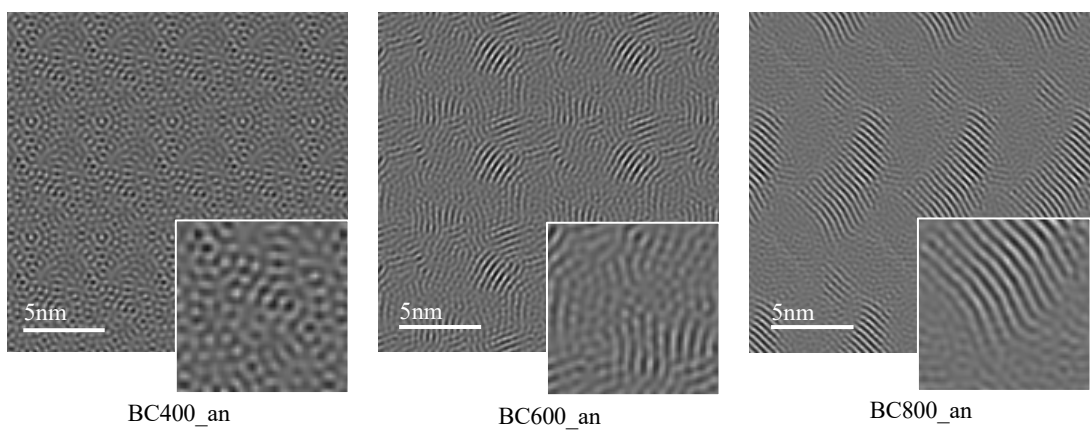

**Figure S5. Simulated TEMs of BC400_an, BC600_an and BC800_an. Insets are 5x5 nm area.**

**S3.3. Atmospheric pressure during annealing**

Throughout the majority of our annealing simulations, we used an applied pressure of 100 bar to encourage the condensation of our systems. Although, this applied pressure may appear large when compared to experimental values, the pressure fluctuations throughout our simulations are on the order of hundreds of bar. Nevertheless, we investigated the effects of this applied pressure during the condensation and equilibration parts of our simulations. To do this, we repeated our simulations using atmospheric pressure. Visualisations of the resulting condensed-phase structures – BC400_1bar, BC600_1bar and BC800_1bar – are shown in **Figure S6**, and their properties are given in **Table S6**.

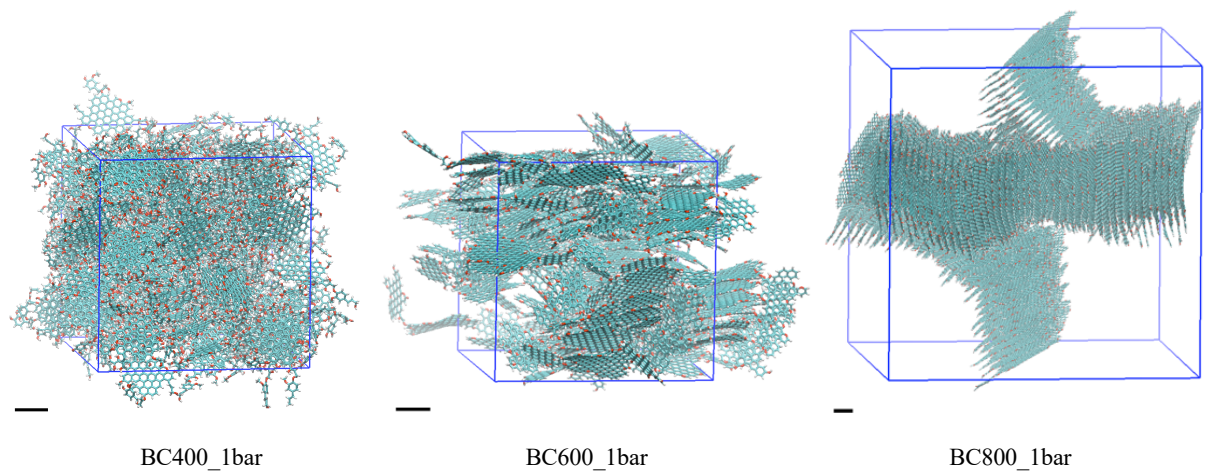

| | BC400_1bar | BC600_1bar | BC800_1bar |

**Figure S6. Visualisations of BC400_1bar, BC600_a1bar and BC800_1bar. Carbon atoms are shown in cyan, oxygen atoms are shown in red and hydrogen atoms are shown in white. Scale bars are 1nm.**

**Table S6. Properties of BC400_1bar, BC600_1bar and BC800_1bar.**

| Model name | H/C | O/C | Aromaticity index /% | *core* aromatic domain size /rings | True density /kg m$^{-3}$ |
|---|---|---|---|---|---|
| BC400_1bar | 0.63 | 0.19 | 78 | 23 | 1397 |
| BC600_1bar | 0.24 | 0.07 | 98 | 75 | 1589 |
| BC800_1bar | 0.09 | 0.04 | 100 | 285 | 1810 |

We found that the *true densities* of the resulting structures were comparable to their high-pressure counterparts, again suggesting *true density* is primarily determined by molecular composition. Visualisations and Simulated TEMs of our systems (**Figure S6 and S7**, respectively), however, revealed that the ordering of our structures increased when formed at lower pressures. This increased ordering is likely due to the more gradual compression and condensation of these systems, allowing the molecular *building blocks* within them to rearrange into more ordered states. We found that, in the BC800 system, our molecular *building blocks* condensed into an infinite needle-like crystal (**Figure S6**) spanning the entire simulation box, even at the high temperatures used in the simulation. Arguably, formation of such structures within the biochar matrix would never be possible during the pyrolysis process and so, we should not be aiming to attain these arrangements. Furthermore, simulation of systems like BC800_1bar, where a small part of the simulation box is occupied by densely packed molecules whilst the remainder is largely unoccupied, is extremely computationally expensive, with poor load management and task distribution. This results in extremely slow computing performance. For these reasons, we chose to proceed with applied pressures of 100 bar during the condensation and equilibration of all systems.

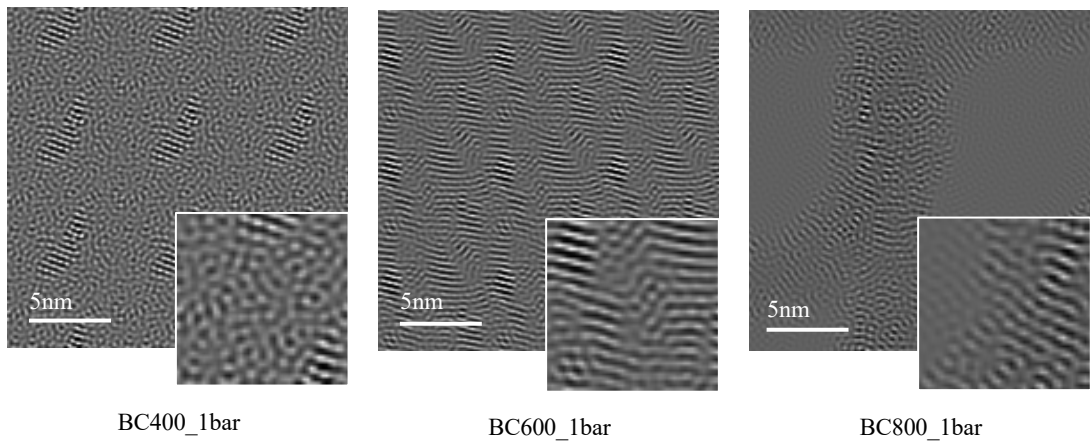

BC400_1bar  BC600_1bar  BC800_1bar

**Figure S7. Simulated TEMs of BC400_1bar, BC600_1bar and BC800_1bar. Insets are 5x5 nm area.**

# Biochars at the molecular level.

# Part 2 – Development of realistic molecular models of biochars.

Rosie Wood[a], Ondřej Mašek[b], Valentina Erastova[a]*

[a] School of Chemistry, University of Edinburgh, Joseph Black Building, David Brewster Road, King's Buildings, Edinburgh, EH9 3FJ, United Kingdom

[b] UK Biochar Research Centre, School of GeoSciences, University of Edinburgh, Crew Building, Alexander Crum Brown Road, King's Buildings, Edinburgh, EH9 3FF, United Kingdom

*valentina.erastova@ed.ac.uk

# Supplementary Information 2

**Assembly of experimental HRTEM images, available in the literature**

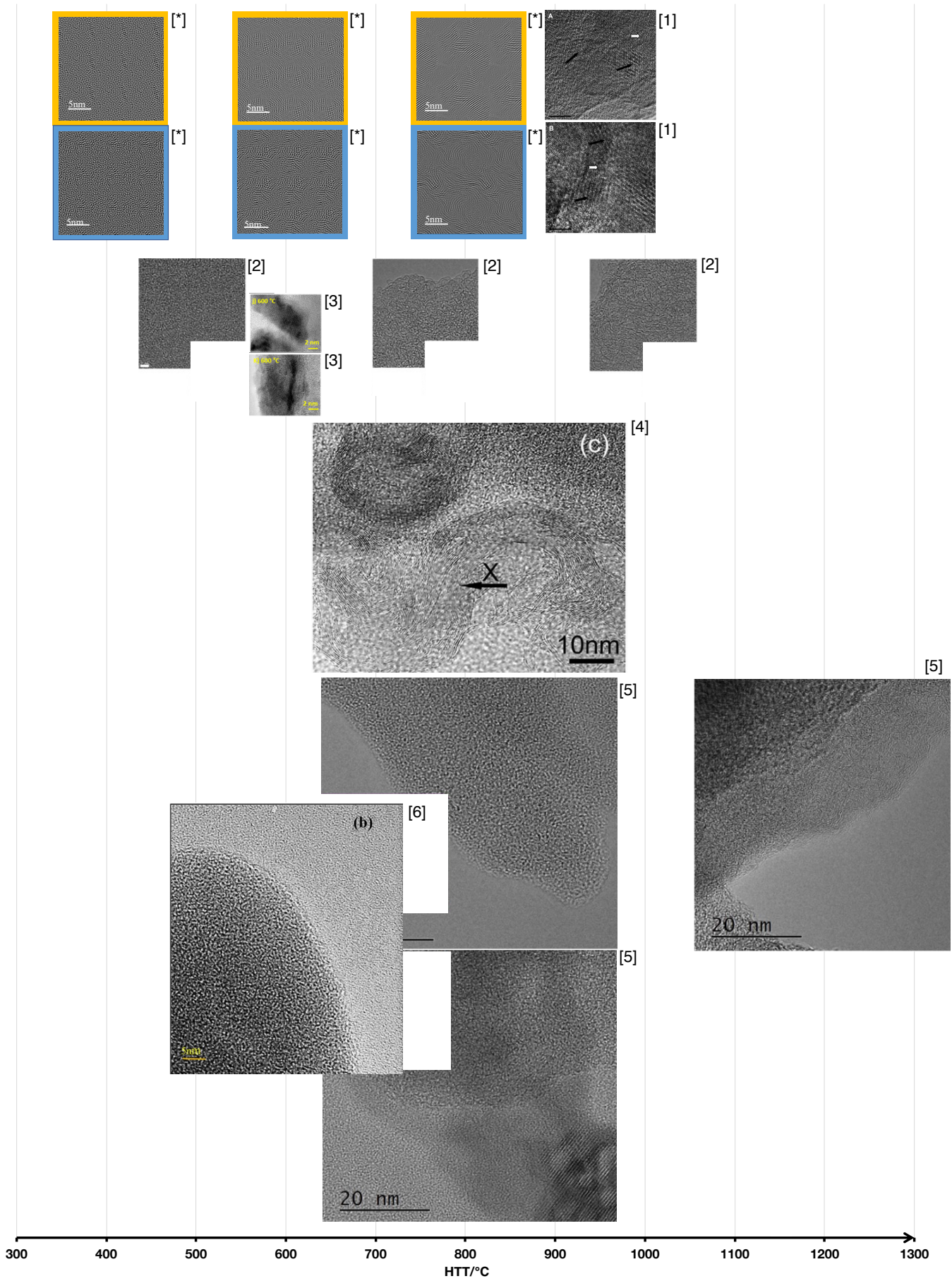

**Figure S8.** A comparative assembly of TEM simulated in this work (orange box highlighting graphitic-only structures, referred to as BC400, BC600 and BC800 in the main text; blue box highlighting the curved structures, BC400c, BC600c and BC800c) and experimental HRTEM from the literature (see referenced below) for biochars produced at a range of HHTs, the images are resized to a common scale.